\documentclass{article}
\usepackage[T1]{fontenc}
\usepackage[utf8]{inputenc}
\usepackage{ismir} %
\usepackage{amsmath,cite,url}
\usepackage{graphicx}
\usepackage{color}
\usepackage{booktabs}
\usepackage{multirow}

\title{Measuring Cross-Cultural Style Diffusion Through Era Classification: US and Korean Popular Music}

\multauthor
  {Dasol Lee$^1$ \hspace{1cm} Minhee Lee$^1$ \hspace{1cm} Seonguk Ju$^1$}
  {{\bf Daewoong Kim$^1$ \hspace{1cm} Harin Lee$^2$ \hspace{1cm} Dasaem Jeong$^1$}\\
  $^1$ Music \& Arts Learning (MALer) Lab, Sogang University, South Korea\\
  $^2$ University of Cambridge\\
  {\tt\small {dasollee, dasaemj}@sogang.ac.kr}
  }

\def\authorname{D. Lee, M. Lee, S. Ju, D. Kim, H. Lee, and D. Jeong}

\usepackage[bookmarks=false,pdfauthor={\authorname},pdfsubject={\pdfsubject},hidelinks]{hyperref}

\begin{document}

\maketitle

\begin{abstract}
Popular music circulates globally while being locally reinterpreted, yet this process of cross-cultural style diffusion has rarely been quantified. We propose an era-classification framework for measuring temporal alignment between chart cultures. CNN classifiers trained from scratch on Billboard Hot 100 audio are applied to Korean Melon chart songs. Korean chart songs from the 1960s through the 1980s are consistently inferred as belonging to earlier Billboard eras, by a median of about four to five years, while the same models remain unbiased on held-out Billboard audio. The offset then halves at the 1990s, to roughly two to three years, and holds there through the 2000s. Reverse inference shows a complementary narrowing, and the pattern holds across architectures and seeds. We interpret these results as reflecting how globally circulating pop styles were locally adopted and progressively synchronized. The framework can be applied to other pairs of chart cultures beyond the US–Korea case examined here.
\end{abstract}

\section{Introduction}\label{sec:introduction}

\begin{figure}[t]
  \centering
  \includegraphics[alt={Schematic overview of the cross-cultural era classification framework. A CNN model trained on Billboard Hot 100 audio is applied to Melon chart audio to measure era offset.},width=\linewidth]{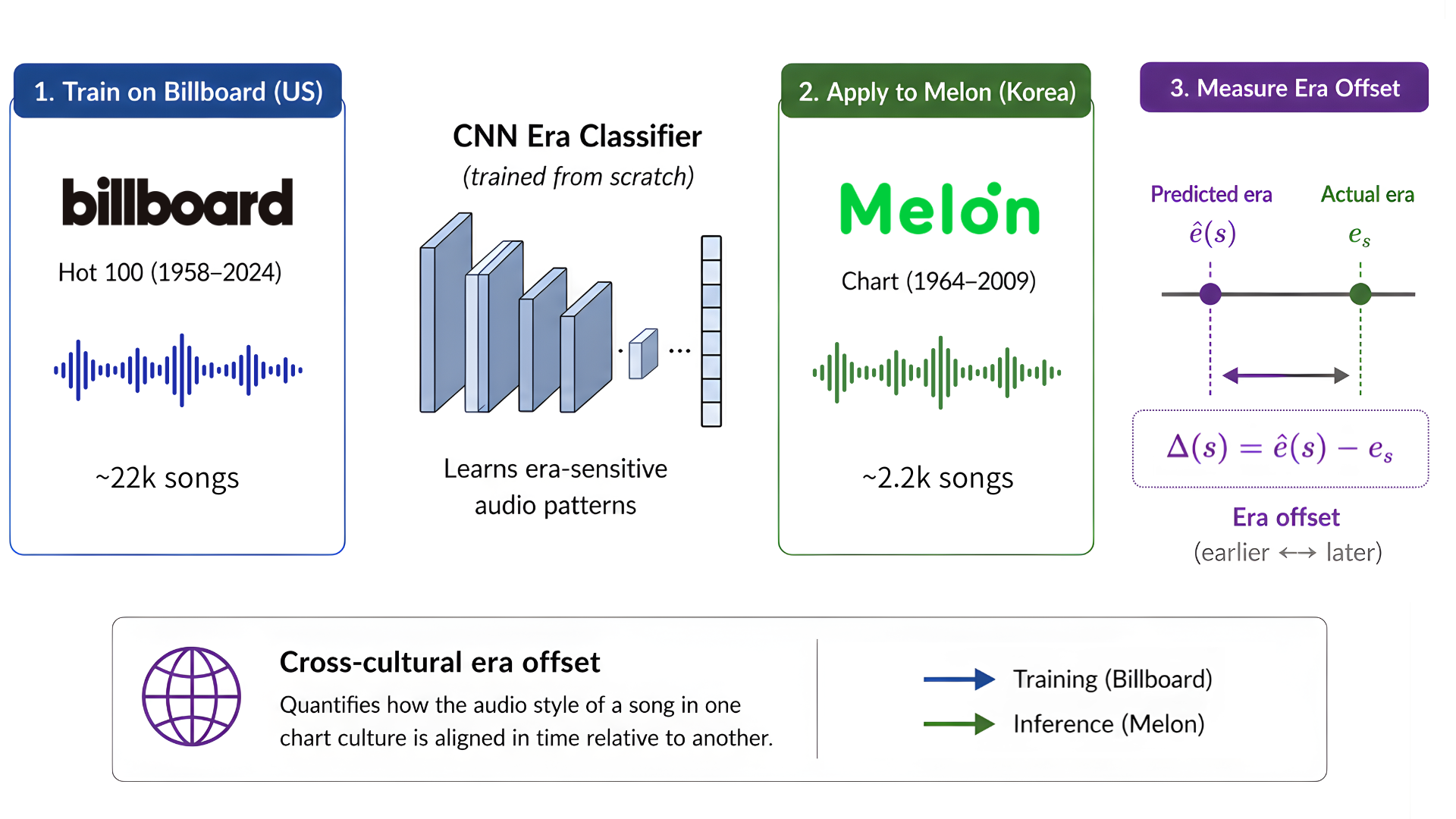}
  \caption{Schematic overview: a CNN era classifier trained on Billboard audio is applied to Melon chart songs. The difference between inferred and actual chart-entry eras quantifies cross-cultural temporal alignment.}
  \label{fig:overview}
\end{figure}

Popular music is a living mirror of its time, changing continuously as musical practices interact with technological conditions, media infrastructures, and transnational cultural flows. Because the modern commercial pop industry first established its large-scale paradigm in the US and UK and later circulated globally \cite{regev2013pop, verboord2016globalization}, many regional pop traditions developed through sustained encounters with these internationally dominant forms while retaining local musical identities.

South Korean popular music is a particularly instructive case. Its
trajectory is extensively documented in musicological literature
(Section~\ref{sec:background}), from the absorption of Western styles
through US military performance circuits in the 1960s to the globally
dominant K-pop system of the 2000s. Despite these rich qualitative
narratives, the acoustic dimensions of this history have been examined
less frequently in quantitative terms.

This paper examines that trajectory computationally. We ask whether audio characteristics that a model associates with particular Billboard eras can also be found in Korean chart music, and if so, with what temporal relationship. We refer to this phenomenon as \emph{cross-cultural style diffusion}, operationalized here as the temporal alignment of era-sensitive audio patterns between two chart ecosystems. This framing presupposes no direct causal influence; it measures the degree to which the two occupy similar positions on a learned temporal axis.

Our core idea is illustrated in \figref{fig:overview}. We train an era classifier exclusively on Billboard Hot 100 audio, using CNN architectures trained from scratch to avoid data leakage from pre-trained models whose corpora likely include Korean music, and apply it to Melon chart songs. If the model systematically infers an earlier Billboard era for a Melon song than its actual chart-entry year, this discrepancy is measured as a negative \emph{era offset}, suggesting that the song's audio characteristics are more similar to those of older Billboard music than to its contemporaries.

Using hierarchical era classification, we find quantitative evidence consistent with long-discussed historical patterns. Korean chart songs first appearing before the 1990s are consistently inferred as belonging to earlier Billboard eras, by a median of about four to five years. This offset then contracts to roughly two to three years in the 1990s and remains at that level through the 2000s.

The contributions of this paper are as follows:
\begin{itemize}
\item We propose an era-based cross-cultural inference framework that uses chronological time---a reference defined consistently for both cultures, and independently of culturally variable genre or mood taxonomies---for quantifying temporal alignment between chart ecosystems.
\item We construct large-scale Billboard (${\sim}$22k tracks) and Melon (${\sim}$2.2k tracks) audio datasets with artist-aware splits designed for cross-domain era inference.
\item We provide empirical evidence that the cross-domain era offset between US and Korean chart music halves at the 1990s and holds thereafter, a pattern that replicates across architectures and seeds and is validated against in-domain predictions from the same models.
\end{itemize}

\section{Background and Related Work}\label{sec:background}

\subsection{Historical Context of Korean Popular Music}

Following the Korean War, the US 8th Army bases served as a major conduit through which Western popular music entered South Korea \cite{jangmijung2022, kimjungha2012}. In the 1960s and 1970s, figures such as \textit{Shin Joong-hyun} and the band \textit{Sanullim} introduced folk and psychedelic rock \cite{HamchunhoandChotaesun2016, jangmijung2019, progrock2023}, drawing on Western formats such as blues-derived structures and electric instrumentation \cite{MMoist2018, library1163} while infusing them with local aesthetics \cite{jangmijung2022, ART002147383}. The debut of \textit{Seo Taiji and Boys} in 1992 then marked the large-scale entry of modern dance music and rap into the Korean mainstream \cite{jangjina2014, seongkiwan2000}; aided by rapid economic growth and changes in media technology \cite{leechorong2020}, the industry moved toward the contemporary K-pop system of the 2000s \cite{K-PopSuwan}. Section~\ref{sec:context} revisits this trajectory against our measurements.

\subsection{Computational Analysis of Musical Evolution}

Long-term change in popular music has been examined computationally using large audio corpora. Serr\`{a} et al.\ \cite{serra2012measuring} revealed statistical trends in pitch, timbre, and loudness across several decades of the Million Song Dataset \cite{bertin2011million}; Mauch et al.\ \cite{mauch2015evolution} tracked the rise and fall of stylistic clusters in the Billboard Hot~100; and Interiano et al.\ \cite{interiano2018musical} documented long-term shifts in audio attributes across over 500{,}000 UK releases. These studies demonstrate that era-sensitive acoustic signatures can be captured at scale, but they primarily describe feature-level trends within a single cultural context.

Parallel to these feature-level analyses, predicting the exact release year or decade of a track has become a standard MIR task \cite{bertin2011million, choi2016automatic}. While typically treated as an end-task for metadata generation, a complementary line of work uses classifiers themselves as analytical lenses. For instance, Nie \cite{nie2022genre} trained genre classifiers on different temporal cohorts of songs from a Chinese music platform and showed that fluctuations in accuracy can signal stylistic evolution rather than algorithmic failure. Our framework shares this insight of reading classification output as a cultural signal, but extends it from within-culture genre evolution to cross-cultural temporal alignment via era prediction.

\subsection{Cross-Cultural MIR}

Cross-cultural MIR research has begun to compare musical corpora across countries and traditions. Liew et al.\ \cite{liew2022network} used chart co-occurrence networks from 30 countries to quantify the cross-cultural reach of Anglo-American popular music, while Papaioannou et al.\ \cite{papaioannou2023west} examined cross-cultural transfer learning across regional datasets, suggesting that learned audio representations carry transferable structural information. These studies operate on a spatial axis by comparing countries at a given time, whereas our work adds a temporal axis by asking \emph{when} on one culture's timeline another culture's songs are positioned.

A key design choice in cross-cultural comparison is the label space. However, cross-cultural label spaces are inherently problematic. Empirical studies show that genre taxonomies and mood ratings exhibit substantial disagreement across different cultural listener populations, and existing taxonomies often carry Western biases \cite{lee2017kpop, hu2012cross, lee2021cross, lee2014cross, celen2025globalmood}. Consequently, chronological time, defined independently of cultural context, provides a comparative axis that is less dependent on culturally variable labels. We therefore use era classification as a comparative lens: rather than asking whether two cultures share genre or mood labels, we ask where songs from one culture fall on a timeline learned from the other.

\section{Datasets}\label{sec:dataset}

\subsection{Billboard Dataset}

We used the Billboard Hot~100 charts (1958--March 2024), retaining only the initial chart entry per song to prevent repeated exposure from periodic re-entries (e.g., holiday songs), yielding 31{,}092 unique entries. To maintain an uncontaminated US pop domain for era-classifier training, we removed 68 tracks primarily credited to Korean acts (e.g., Wonder Girls, BTS), as they exhibit distinct characteristics of contemporary K-pop. Secondary features by Korean artists were retained.

For audio collection, we queried YouTube and filtered out non-original versions (e.g., live, cover, music videos) using keyword heuristics and fuzzy string matching on metadata. Through this pipeline, we obtained audio for 22{,}002 tracks. 

\subsection{Melon Dataset}

For the Korean dataset, we used the Melon chart, which is uniquely suited to this study as it is the only major chart aggregating historical rankings since the 1960s. Its format evolved from annual lists (1964--1984) to weekly formats thereafter. Using a pipeline analogous to that for Billboard, we collected audio for approximately 2{,}200 tracks that entered the chart between 1964 and 2009. 

We intentionally capped the Melon timeline at 2009 to observe the historical period of unidirectional style adoption. Because the full-scale globalization of K-pop in the 2010s led to bidirectional cultural flows \cite{K-PopSuwan, miroudot2024kpop}, restricting our timeline to pre-2010 keeps the measured era offset interpretable as local charts converging toward global trends rather than mixing with them.

For both datasets, era labels are assigned based on the date of first chart entry rather than the release date. First chart entry more directly reflects when a song entered collective listening and commercial circulation, which is particularly relevant for studying style diffusion across chart ecosystems. This choice means that re-entries, posthumous re-releases, and remakes receive labels reflecting their chart circulation date rather than their original production date. 

\subsection{Artist-Aware Split and Balancing}

\begin{table}[t]
  \centering
  \footnotesize %
  \setlength{\tabcolsep}{3.5pt} %
  \begin{tabular}{@{}llcccccc@{}}
    \toprule
    Dataset & Split & 1960s & 1970s & 1980s & 1990s & 2000s & 2010s \\
    \midrule
    \multirow{3}{*}{Billboard}
      & Train & 4769 & 2850 & 1970 & 1600 & 1585 & 4387 \\
      & Val.  & 1005 &  555 &  294 &  279 &  134 &  153 \\
      & Test  &  987 &  484 &  404 &  191 &  123 &  232 \\
    \midrule
    \multirow{3}{*}{Melon}
      & Train &   34 &  191 &  410 &  494 &  533 &    0 \\
      & Val.  &   18 &   23 &   48 &   94 &  121 &    0 \\
      & Test  &   14 &   37 &   44 &   85 &  125 &    0 \\
    \bottomrule
  \end{tabular}
  \caption{Tracks per decade in each split. Songs from 1958--1959 are
  grouped with the 1960s, 2020--2024 with the 2010s.}
  \label{tab:dataset_split}
\end{table}

To prevent artist leakage (i.e., the model recognizing a singer's voice rather than era-sensitive musical patterns), we split both datasets by artist. Because collaborations create implicit links between artists, we constructed a collaboration graph in which two artists are connected if they share at least one chart entry, and identified its connected components. Each connected component was assigned as a whole to train, validation, or test with an approximate ratio of 8:1:1. Artists with no collaborations formed singleton components. The resulting split sizes are shown in Table~\ref{tab:dataset_split}.

To mitigate class imbalance across eras, we applied class-balanced undersampling at the decade level during training. We note that the Melon chart's shift from annual to weekly format means that earlier decades (1960s--1970s) contain substantially fewer tracks than later ones; this should be kept in mind when interpreting per-decade results for those periods.

\section{Method}\label{sec:method}
We formulate the measurement of cross-cultural temporal alignment as \textit{cross-domain era classification}. A classifier trained to predict chart-entry eras in one culture is applied to songs from another; the systematic discrepancy between predicted and actual eras quantifies the temporal offset.

Concretely, let $s$ be a song whose actual chart-entry era is $e_s$. When a Billboard-trained model infers an era $\hat{e}(s)$ for a Melon song, the \emph{era offset} is \begin{equation}\label{eq:era_offset} \Delta(s) \;=\; \hat{e}(s) \;-\; e_s\,. \end{equation}
A negative mean offset indicates that those songs are positioned earlier on the Billboard-derived temporal axis than their actual chart-entry dates would suggest. At inference time, each track is segmented into non-overlapping 30-second windows, their year-level softmax outputs are averaged, and $\hat{e}(s)$ is the year holding the largest averaged probability. Decade-level summaries are obtained via Gaussian kernel density estimation (bandwidth${}=1$\,year); we report the mode of each distribution.

\subsection{Hierarchical Era Prediction}\label{sec:hierarchy}

Treating year-level prediction as a flat multi-class problem ignores the ordinal structure of time. We therefore train the model to jointly predict the era at four hierarchical levels: decade, half-decade, quarter-decade, and year. Since the Billboard dataset spans 1958--2024, songs from 1958--1959 are assigned to the 1960s and songs from 2020--2024 to the 2010s. To maintain a consistent tree in which every child node is wholly contained within its parent, the third level divides each half-decade into intervals of 2 or 3 years rather than exactly 2.5, ensuring that no year maps to two different third-level nodes.

We prefer this to direct year regression because a squared-error objective would smooth away the songs of most interest here. Some tracks carry traits of two separated eras---12\,\% of Melon tracks receive a year distribution with two well-separated peaks---and a classifier keeps both, whereas regression would answer with the midpoint, a year resembling neither.

Each hierarchical level has an independent softmax head, and the cross-entropy loss is summed over levels: $\mathcal{L}_{CE} = \sum_{\ell=0}^{3} \mathcal{L}_{CE}^{(\ell)}$, where $\ell$ indexes the four levels from decade to year. Because the heads predict independently, their outputs may be mutually inconsistent, so we add a hierarchical consistency loss $\mathcal{L}_{HC}$~\cite{wehrmann2018hierarchical} penalizing parent--child violations, giving $\mathcal{L}_{\text{total}} = \mathcal{L}_{CE} + \lambda\mathcal{L}_{HC}$ with $\lambda{=}1$ throughout. 

\subsection{Audio Representation and Training Protocol}
All audio was converted to mono at 16\,kHz, deliberately discarding the stereo image and everything above 8\,kHz, where mastering differences are most audible. Spectrograms follow each architecture's original settings.\footnote{Code, chart-entry labels with YouTube IDs, artist-aware splits, and full training configurations: \url{https://github.com/malerlab/billboard-melon-era}} During training, a random 30-second crop is drawn from each track per epoch.

We trained all models with the Adam optimizer (lr\,$=$\,$1\!\times\!10^{-4}$) and a batch size of~64 for 30\,000 iterations, validating every 500, and retained the checkpoint with the highest validation macro accuracy. To mitigate class imbalance, training batches were constructed by undersampling each decade to match the size of the smallest class (2000s, 1\,585 tracks). This procedure is re-randomized every epoch. Every architecture was trained under three random seeds, giving the 18 runs over which we report variability below.

\subsection{Model Selection Strategy}\label{sec:model_selection}

A central design requirement is that the era classifier must be trained exclusively on audio from a single chart culture so that its temporal representations are derived from that culture alone.
Large-scale pre-trained music foundation models such as MERT or Jukebox are trained on internet-scale corpora that very likely include Korean popular music; fine-tuning such models would risk cross-cultural data leakage, confounding the measurement we aim to perform.

We therefore use CNN architectures trained from scratch. We evaluated five models commonly used in MIR~\cite{won2020eval}---FCN
\cite{choi2016automatic}, ShortChunkCNN and its residual variant (ShortChunkCNN\_Res)~\cite{won2020data}, musicnn~\cite{pons2017end}, and CRNN~\cite{choi2017convolutional}---plus a simple baseline of three 2D convolutional layers (kernel size~3), each followed by batch normalization, ReLU, and max pooling.
Model sizes range from approximately 86\,k (baseline) to 837\,k parameters. The goal is not to maximize era-prediction accuracy per se, but to ensure that whatever temporal patterns the models learn are attributable solely to Billboard audio, making cross-domain inference interpretable as a culturally uncontaminated lens.

\section{Results}\label{sec:results}

\subsection{In-Domain Performance on Billboard}

\begin{table}[t]
  \centering
  \small
  \begin{tabular}{@{}l*{6}{c}@{}}
    \toprule
    Decade & CNN & FCN & SCNN & SCNNR & Musicnn & CRNN \\
    \midrule
    60s & 90.8 & 84.3 & 89.4 & 83.7 & 84.6 & 81.1 \\
    70s & 66.3 & 84.6 & 69.8 & 80.1 & 66.9 & 70.6 \\
    80s & 78.0 & 68.8 & 85.9 & 73.3 & 83.7 & 88.4 \\
    90s & 56.8 & 64.6 & 68.8 & 60.4 & 61.5 & 57.8 \\
    00s & 74.8 & 85.4 & 79.7 & 72.4 & 73.2 & 87.0 \\
    10s & 35.3 & 30.2 & 33.6 & 38.8 & 50.9 & 28.9 \\
    \midrule
    Macro & 67.0 & 69.6 & 71.2 & 68.1 & 70.1 & 69.0 \\
    Micro & 75.0 & 75.1 & 77.5 & 74.6 & 75.3 & 73.7 \\
    \bottomrule
  \end{tabular}
  \caption{Track-level era classification accuracy (\%) on the Billboard test set at the decade level, for the seed-77 run of each architecture. Across all 18 runs, macro accuracy is $69.0{\pm}2.0$ and micro accuracy $75.3{\pm}1.5$.}
  \label{tab:modelwise_acc}
\end{table}

\begin{figure}[t]
  \centering
  \includegraphics[alt={Confusion matrix of year-level era classification on the Billboard test set by the baseline CNN model. A clear diagonal tendency is visible, indicating that predictions are close to the true era even when not exact. The tendency weakens around the 2010s.},width=\linewidth]{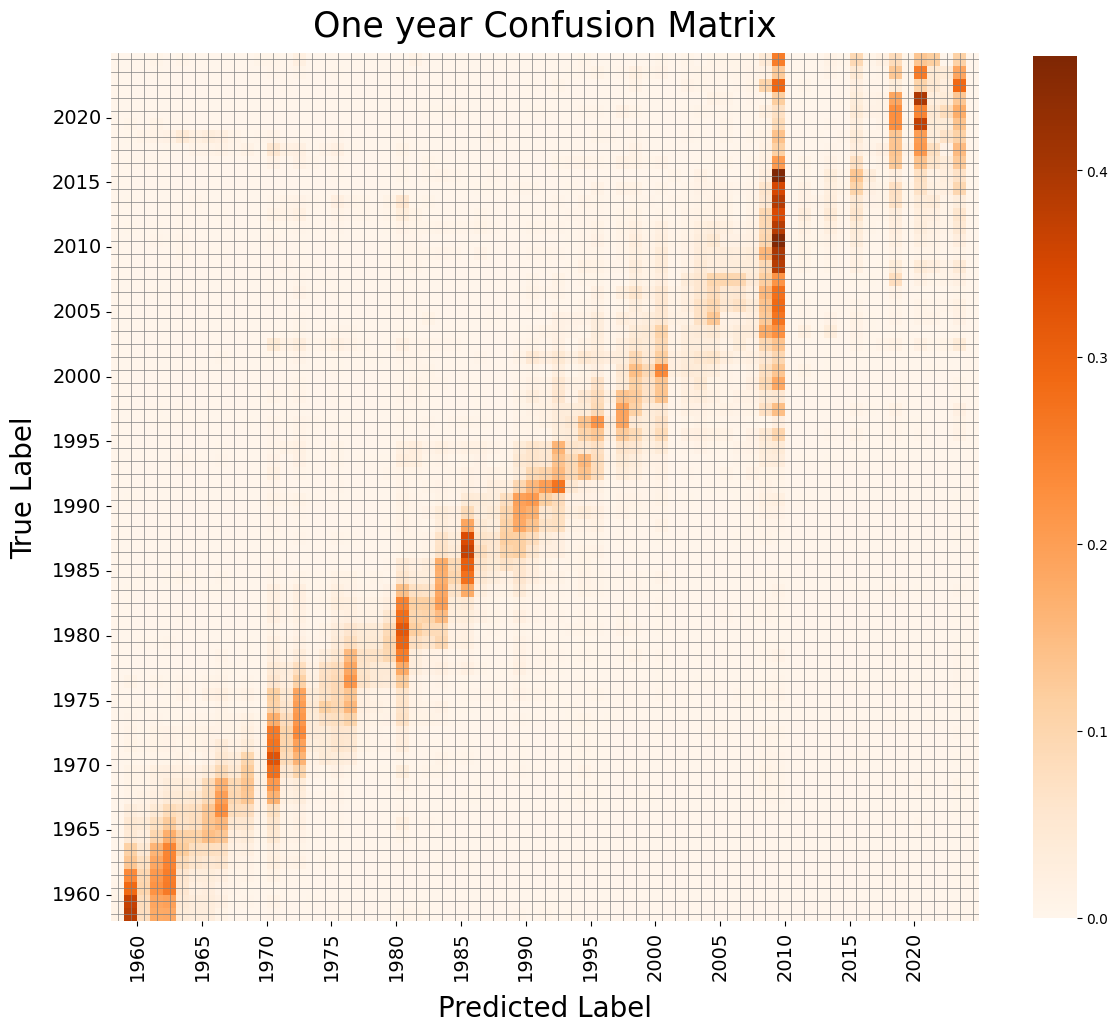}
  \caption{Confusion matrix of year-level predictions on the Billboard test set, pooled over all runs. Even when the exact year is missed, predictions fall near the true era, as the diagonal shows. The tendency weakens around the 2010s, possibly reflecting greater stylistic diversity in recent music.}
  \label{fig:confusion}
\end{figure}

All six architectures achieved reasonable in-domain performance on the
Billboard test set (\tabref{tab:modelwise_acc}), with decade-level
macro accuracy ranging from 67.0\,\% to 71.2\,\%, well above the
16.7\,\% chance level for six-class classification. The year-level
confusion matrix (\figref{fig:confusion}) further shows that errors
are not random: predictions cluster near the diagonal, indicating that
the models capture a meaningful temporal gradient. This tendency
weakens around the 2010s, which may reflect the rapid diversification
of genres and production techniques in recent popular music.

\subsection{Billboard-to-Melon Inference: Main Finding}

\begin{figure}[t]
  \centering
  \includegraphics[alt={Bar plot showing the prediction error distribution for Billboard and Melon datasets by quarter decade. Billboard errors center around zero while Melon errors are biased toward earlier eras.},width=0.80\linewidth]{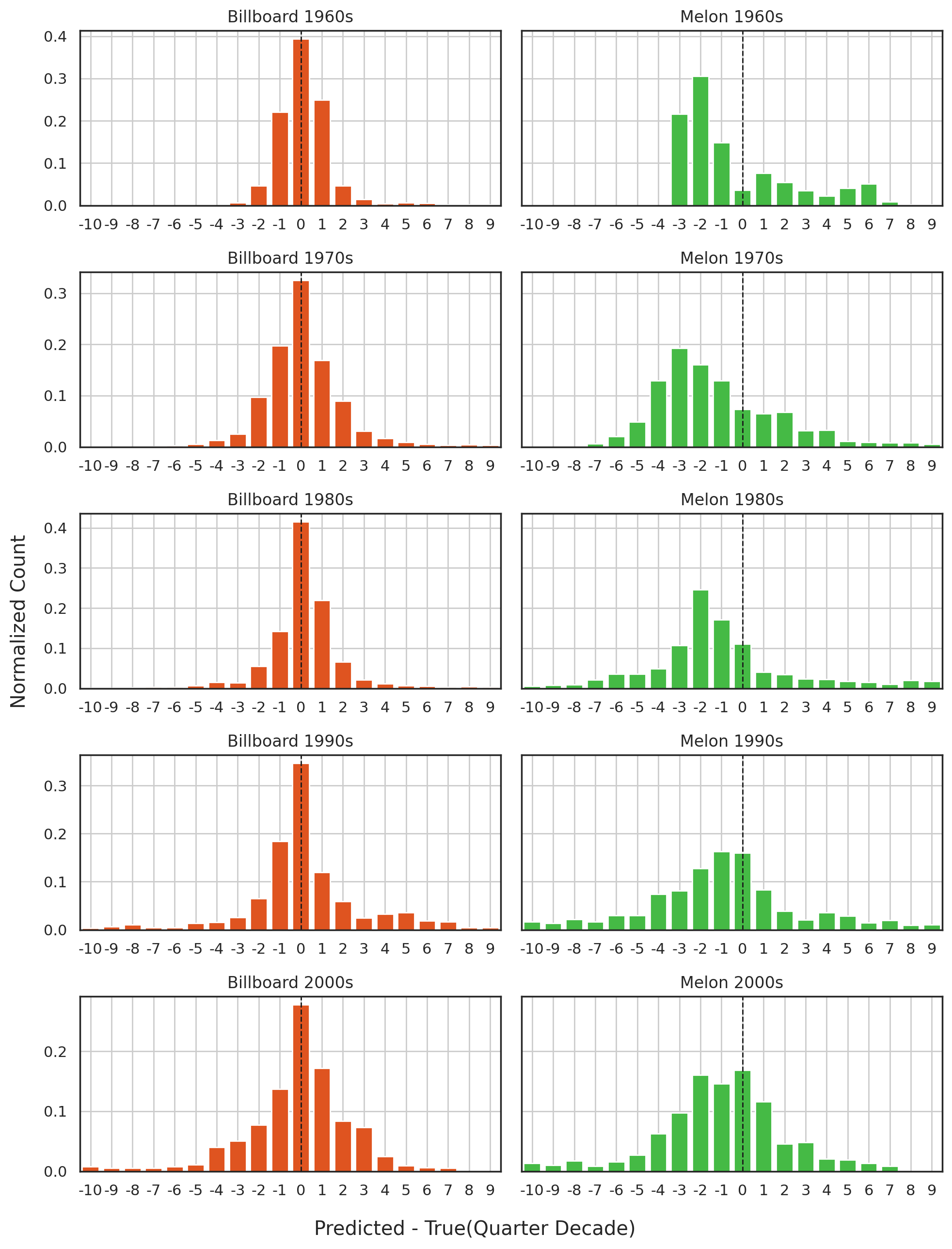}
  \caption{Prediction error distribution (quarter-decade level) for Billboard (in-domain) and Melon (cross-domain), pooled over all runs, with the $y$-axis shared within each decade so the two domains are directly comparable. Billboard errors stay centered on zero in every decade while widening toward the 2000s; Melon errors are displaced toward earlier eras throughout.}
  \label{fig:barplot}
\end{figure}

\begin{figure}[t]
  \centering
  \includegraphics[alt={KDE plot showing year-level prediction results for Melon songs by decade when using a Billboard-trained model. The era gap is pronounced in the 1960s to 1980s and contracts from the 1990s onward.},width=\linewidth]{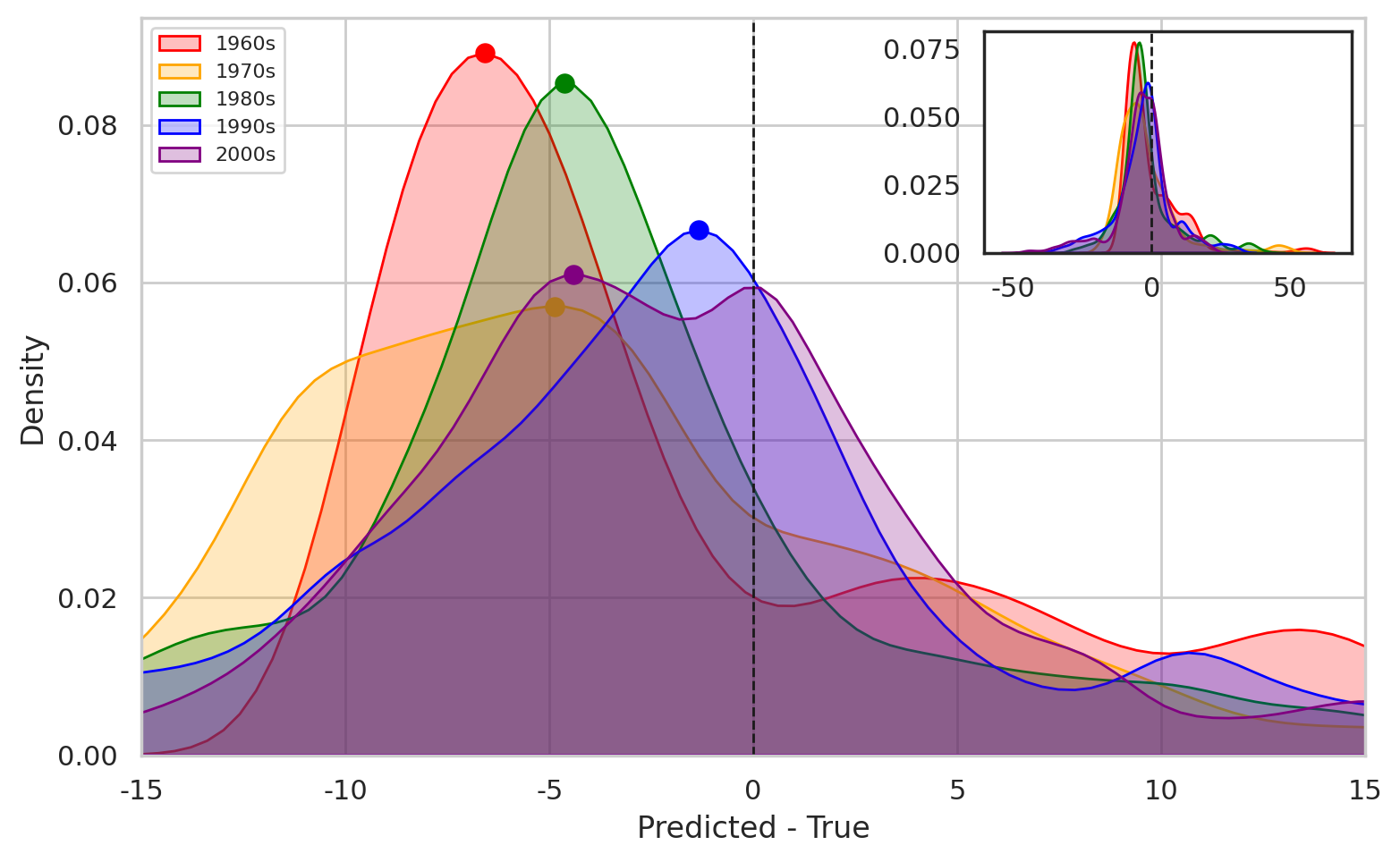}
  \caption{KDE of year-level predictions for Melon songs (Billboard $\rightarrow$ Melon), pooled over all runs. Every decade peaks left of zero. The 2000s curve is the only bimodal one, with mass both at the origin and near $-4$ years: part of the decade has synchronized and part has not. Inset: the same densities on the full prediction range.}
  \label{fig:kde}
\end{figure}

\begin{table}[t]
  \centering
  \small
  \setlength{\tabcolsep}{3.2pt}
  \begin{tabular}{@{}lccccc@{}}
    \toprule
     & 1960s & 1970s & 1980s & 1990s & 2000s \\
    \midrule
    Melon tracks ($n$) & 166 & 351 & 602 & 773 & 879 \\
    Median offset (yr) & $-4.7$ & $-4.5$ & $-4.1$ & $-2.4$ & $-2.7$ \\
    \phantom{Median offset (yr)}\,${\pm}$SD & $0.6$ & $0.7$ & $0.8$ & $0.8$ & $1.4$ \\
    \bottomrule
  \end{tabular}
  \caption{Median era offset per decade for Melon songs under
    Billboard-trained models, mean ${\pm}$ SD across the 18 runs. The
    same models are unbiased in domain (\figref{fig:barplot}).}
  \label{tab:era_offset}
\end{table}

When Billboard-trained models are applied to Melon songs, the
prediction error is no longer centered around zero
(\figref{fig:barplot}), while on held-out Billboard audio the same
models stay centered in every decade, with a median in-domain offset of
at most $0.2$ years. Melon songs are consistently
inferred as belonging to earlier Billboard eras.

\tabref{tab:era_offset} summarises the median era offset for each
decade. The offset holds near $-4.7$, $-4.5$, and $-4.1$ years across
the 1960s, 1970s, and 1980s. The 1960s value is bounded from below
because the model cannot predict eras earlier than 1958, and the Melon
corpus begins in 1964: its earliest songs can be shifted at most six
years into the past. In the 1990s the offset contracts to $-2.4$
years, a step reproduced in every run, and holds at that level through
the 2000s ($-2.7$, the least stable value in the table and the decade
where in-domain errors are most dispersed as well,
\figref{fig:barplot}).
This pattern is consistent with musicological accounts suggesting that
Korean popular music exhibited a temporal offset in adopting globally
circulating styles in earlier decades, and that this offset narrowed
sharply after the 1990s transition without closing.

In the 2000s the predictions also split in two
(\figref{fig:kde}): some songs are dated on-era, others still several
years early, and no single number describes both. The split follows
genre. Among the tracks dated on-era in at least
80\,\% of runs, the recurring names are dance and hip-hop acts
(BigBang, 2NE1, Epik High), while those dated consistently early are
led by ballad and R\&B singers (SG Wannabe, Brown Eyes, Wheesung).

\subsection{Cross-Architecture Consistency}

\begin{figure}[t]
  \centering
  \includegraphics[alt={KDE plot of 1970s and 2000s predictions for all six
    CNN architectures, with an arrow per model from its 1970s density mode
    to its 2000s mode. Every arrow points toward zero, by differing amounts.},
    width=\linewidth]{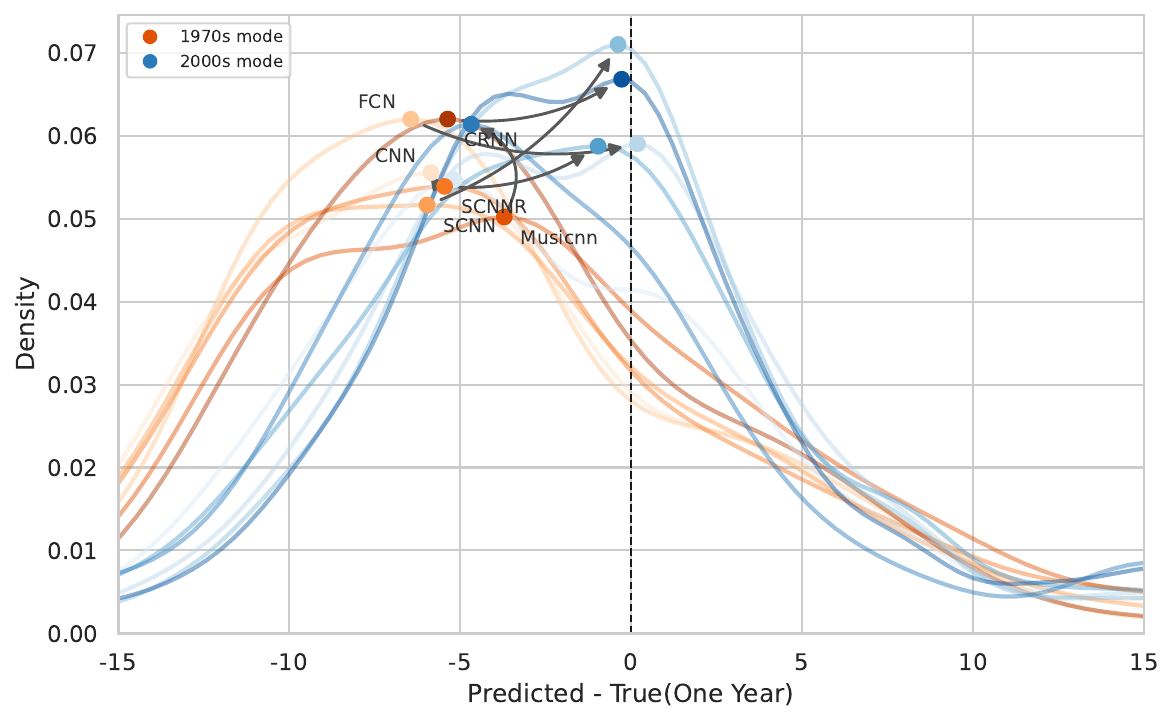}
  \caption{Per-architecture shift of the era offset
    (Billboard $\rightarrow$ Melon), each model pooled over its seeds.
    Arrows run from a model's 1970s density mode (orange) to its 2000s
    mode (blue).}
  \label{fig:kde_models}
\end{figure}

The overall trend is not an artifact of a single architecture.
\figref{fig:kde_models} compares the 1970s and 2000s predictions
across all six models. Every model back-dates the 1970s, by
between $3.7$ and $6.5$ years. Moving to the 2000s, four of the six
shift four to six years toward zero: on most architectures the
distance between the two chart cultures has largely closed over these
three decades. The other two dissent---the baseline CNN barely moves,
and Musicnn moves slightly further back---so how much offset survives
into the 2000s remains architecture-dependent.

\subsection{Melon-to-Billboard Reverse Inference}

\begin{figure}[t]
  \centering
  \includegraphics[alt={KDE plot showing year-level prediction results
    for Billboard songs by decade when using a Melon-trained model.
    Billboard songs are predicted toward later eras on the
    Melon-derived timeline.},
    width=\linewidth]{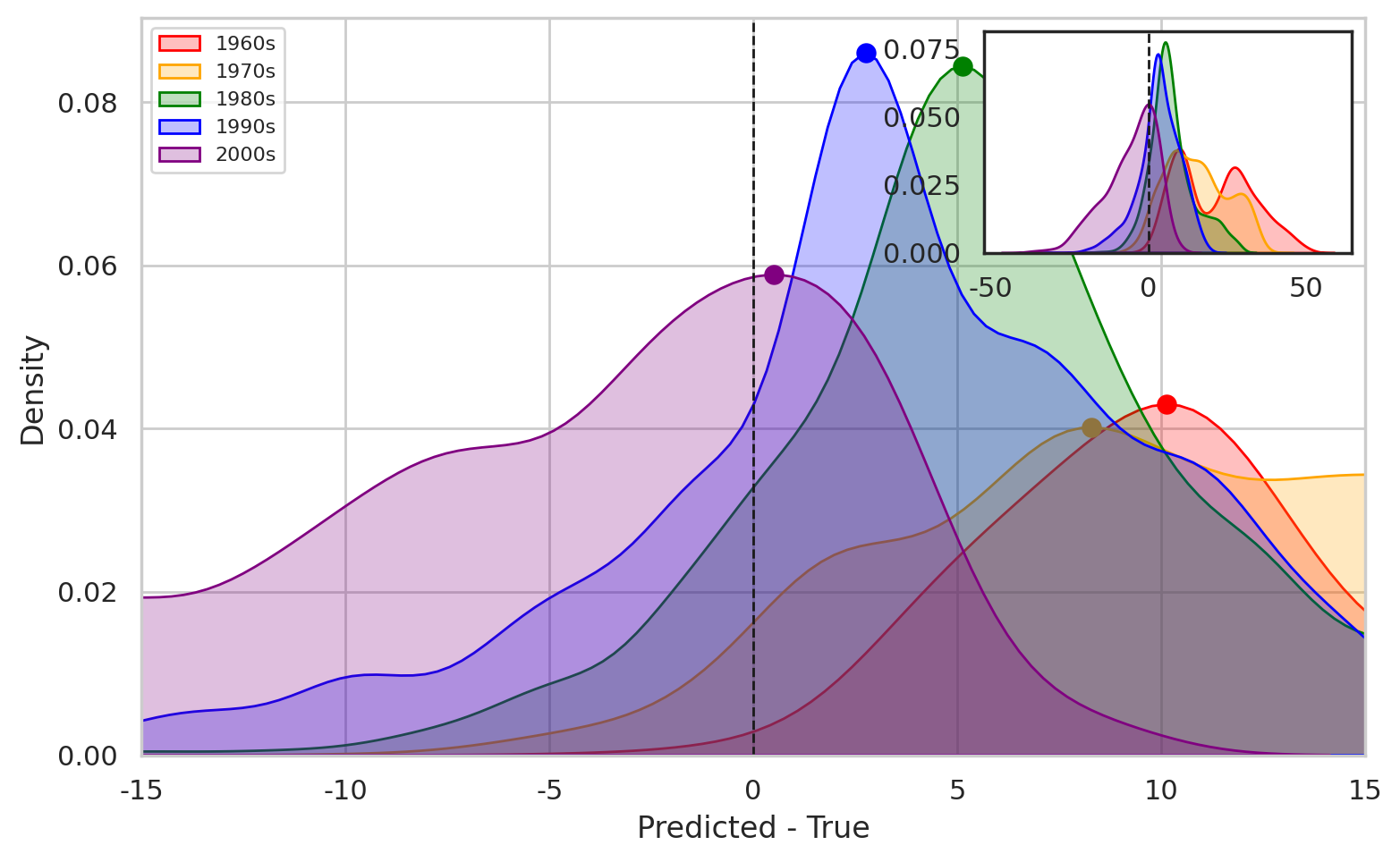}
  \caption{KDE of year-level predictions for Billboard songs
    (Melon $\rightarrow$ Billboard), pooled over seeds of the
    Melon-trained model. Billboard songs are predicted toward later
    eras on the Melon-derived timeline.}
  \label{fig:kde_reverse}
\end{figure}

Back-dating might simply be what a Billboard-trained model does to
unfamiliar audio; reversing the direction tests that. Trained on Melon
and applied to Billboard, the models place Billboard songs later on
the Melon-derived timeline, by amounts shrinking from about ten years
in the 1960s to near zero in the 2000s
(\figref{fig:kde_reverse})---the same narrowing, recovered by models
that never saw Billboard. It carries less weight: the Melon-trained
model reaches 52.4\,\% macro accuracy on the artist-disjoint Melon test
split, against 76.3\,\% for the Billboard-trained models over the same
five decades, and 9.5\,\% for the 1960s, whose training partition holds
34 tracks. We read it as corroborating the
direction of the effect rather than its magnitude.

\section{Contextualizing the Era Offset}\label{sec:context}

\begin{table}[t]
\centering
\small
\setlength{\tabcolsep}{3.5pt}
\begin{tabular}{lcrcc}
\toprule
Artist & Period & $n$ & Median (yr) & Spread \\
\midrule
Choi Hee-jun & 1964--69 & 12 & $-6.1{\pm}0.7$ & $-8$ to $-2$ \\
Sanullim & 1978--88 & 13 & $-5.2{\pm}2.0$ & $-13$ to $+1$ \\
Seo Taiji & 1992--03 & 10 & $+2.8{\pm}2.2$ & $-5$ to $+10$ \\
BigBang & 2007--09 & 14 & $-0.1{\pm}0.8$ & $-5$ to $+1$ \\
Girls' Gen. & 2007--09 & \phantom{0}6 & $-1.0{\pm}1.1$ & $-6$ to $+1$ \\
\bottomrule
\end{tabular}
\caption{Year-level era offsets for five representative Korean artists.
Median is the mean ${\pm}$ SD of the per-artist median across runs;
spread is the 10th--90th percentile of that artist's per-song offsets,
after excluding two Sanullim entries whose audio turned out to be
non-original recordings.}
\label{tab:casestudy}
\end{table}

Table~\ref{tab:casestudy} summarizes the model's predictions 
for five representative Korean artists spanning four decades. 
We interpret these patterns in light of the structural 
conditions of cross-cultural music flow.

\subsection{Structural Delay under Limited Channels 
(1960s--1970s)}

Until the late 1980s, Korean musicians accessed American popular music mainly through U.S.\ military base performances and AFKN radio~\cite{maliangkay2006a, kimpilho2016}. Choi Hee-jun, who has recalled formative listening to artists such as Nat King Cole via military FM~\cite{choiheejun_interview}, and Sanullim, whose psychedelic rock drew on styles available through the same channels~\cite{kimandsin2010}, are back-dated by $6.1$ and $5.2$ years despite a full decade separating their activity. That the gap persists across genres and generations is consistent with its being a property of the distribution infrastructure rather than of individual artists.

\subsection{The Seo Taiji Watershed and Its Aftermath}

Seo Taiji and Boys (debut 1992) are widely credited with dismantling the broadcast-network-controlled Korean music market and catalyzing the rise of the talent-agency system that defines contemporary K-pop~\cite{shim2006, jung2006, maliangkay2013}.
Shim~\cite{shim2006} notes that ``each of their albums was in itself a musical experimentation,'' spanning new-jack-swing, metal with traditional instruments, gangsta rap, and techno.
Seo Taiji is the only artist in Table~\ref{tab:casestudy} the model places \emph{ahead} of his chart dates, by $+2.8$ years and in 15 of 18 runs, at a time when his decade as a whole is still back-dated by $2.4$ years. His songs also spread over a 15-year band, the widest in the table, against six to seven years for the consistently placed artists on either side of the transition.

This is the transition our decade-level results locate in the 1990s,
visible in a single career: by breaking the broadcast monopoly, he
helped open the way for agencies such as SM Entertainment and YG
Entertainment (the latter founded by his former bandmate Yang
Hyun-suk) to produce artists oriented toward global trends from
inception. BigBang (YG) and Girls' Generation (SM) are consistent with
this reading, with median offsets near zero by 2007--09. We stress that
a single watershed figure cannot be shown to have caused a
corpus-level shift; what the measurements show is that the structural
gap contracted sharply around this moment.

\section{Discussion and Conclusions}\label{sec:discussion}

We proposed an era-classification framework to quantify cross-cultural temporal alignment. Our main finding---a four-to-five year past-biased era offset in Korean chart music from the 1960s through the 1980s, halving at the 1990s and holding at two to three years through the 2000s---gives quantitative form to musicological accounts of how globally circulating styles were locally adopted and progressively synchronized.

This study has several limitations. Our models cannot separate compositional style from recording technology: a mel-spectrogram CNN responds to tape noise, compression, and mastering as readily as to harmony or instrumentation, so part of the measured offset is likely production lag rather than stylistic lag. Related factors point the same way: 30-second crops give no access to song-level form, and artist-aware splitting does not control for producers, engineers, or songwriters. Individual songs are dated imprecisely: across the 30-second crops of one song, predicted years vary with a mean SD of 3.3 years on Billboard and 6.0 on Melon, so while our decade medians average that noise out, no one song's offset is a precise dating. The audio is matched from YouTube by title and artist and is not always the original recording. Finally, chart data reflects circulation rather than production, and the corpora cover different spans.

Two of these we can bound. Mastering style alone does not drive the era estimate: across 243 pairs of original and remastered audio for the same songs, the predictions that change divide almost evenly between earlier and later, as the mono 16\,kHz input intends. And restricting the corpus to its best-matched 80\,\% moves no decade median by more than half a year.

We read the era offset not as simple imitation, but as a trace of the evolving structural conditions of cross-cultural music flows. Future work will extend the framework to other chart systems, identify the acoustic cues behind the offset, and apply era inference within a single corpus to surface artists predicted into later eras than their own~\cite{collins2025career}. One such case already sits in our Billboard test set: the models date The Beatles' 1960s entries about two years later than their chart dates, against essentially zero for the other 1960s songs in the split, a gap that holds in 15 of the 18 runs. The Beatles are widely credited with pioneering studio production, and the models place their music accordingly.

\section{Acknowledgment}
This work was supported by the Ministry of Education of the Republic of Korea and the National Research Foundation of Korea (NRF-2024S1A5C3A03046168).

\end{document}